\documentclass[12pt]{article}
\usepackage[margin=1.9cm]{geometry}
\usepackage{epsfig}
\usepackage{cite}

\newcommand{\mysection}{\setcounter{equation}{0}\section}

\def\beq{\begin{equation}}
\def\eeq{\end{equation}}
\def\beqa{\begin{eqnarray}}
\def\eeqa{\end{eqnarray}}
 
\begin{document}

\begin{center}
{\Large \bf Soft-gluon corrections for $tqZ$ production}
\end{center}

\vspace{2mm}

\begin{center}
{\large Nikolaos Kidonakis$^a$ and Nodoka Yamanaka$^{b,c}$}\\

\vspace{2mm}

${}^a${\it Department of Physics, Kennesaw State University, \\
Kennesaw, GA 30144, USA}

\vspace{1mm}

${}^b${\it Kobayashi-Maskawa Institute for the Origin of Particles and the Universe, Nagoya University, \\ Furocho, Chikusa, Aichi 464-8602, Japan}

\vspace{1mm}

${}^c${\it Nishina Center for Accelerator-Based Science, RIKEN, \\ Wako 351-0198, Japan}
\end{center}

\begin{abstract}
We study soft-gluon corrections for the associated production of a single top quark and a $Z$ boson ($tqZ$ production) at hadron colliders. We find that the radiative corrections are dominated by soft-gluon emission. We calculate the approximate NNLO (aNNLO) cross section at LHC energies, including uncertainties from scale dependence and from parton distributions. We also calculate differential distributions in top-quark rapidity. We show that the aNNLO corrections are significant and they enhance the NLO cross section while decreasing theoretical uncertainties.
\end{abstract}

\mysection{Introduction}

The production of a single top quark in association with a $Z$ boson, i.e. $tqZ$ production where $q$ represents a light quark or antiquark, is an interesting process that is being studied at the LHC \cite{CMS1,ATLAS1,CMS2,CMS3,ATLAS2,CMS4}. Searches for the process were already underway at 8 TeV LHC energy \cite{CMS1} with later measurements following at 13 TeV \cite{ATLAS1,CMS2}, and eventually leading to the observation of this process at 13 TeV collisions \cite{CMS3,ATLAS2}. Further measurements, including differential cross sections, were made in \cite{CMS4}.

The $tqZ$ production processes allow for the $t$-$Z$ and $W$-$W$-$Z$ couplings to be studied in a single interaction. These processes can be affected by physics beyond the Standard Model, including flavor-changing neutral-current processes with anomalous top-quark couplings, such as $t$-$q$-$Z$ \cite{NKAB,NKtZ,LZLGZ,MGNK,YLSM} and $t$-$q$-$g$ \cite{NKEM}. Top partner decays to $tZ$ were considered in \cite{JRMT}. Furthermore, analyses for $tqZ$ production in the context of SMEFT were presented in \cite{DMMVZ,RBAI}.

The $tqZ$ processes also probe the anomalous ``weak'' moments of the top quark, whose interactions are defined as 
${\cal L}_{tZ}=C_t \bar t \sigma_{\mu \nu} (\partial^\mu Z^\nu ) t +C'_t \bar t \sigma_{\mu \nu} (\partial^\mu Z^\nu ) \gamma_5 t$. These anomalous moments may be generated by new particles beyond the Standard Model appearing through loop diagrams in a similar way as the anomalous electromagnetic moments. On the other hand, these moments may also be probed by low-energy precision-test experiments. For instance, the top-quark weak dipole moment contributes to the electric dipole moments (EDMs) of light quarks through renormalization group evolution \cite{CDVM}, and finally to the observable EDMs of the neutron and atoms \cite{Yamanaka2017}. However, the mixing between top-quark weak moments and the electromagnetic moments of light quarks occurs through the weak interaction, so a significant suppression happens. We then expect the direct production in accelerator experiments to have an important advantage over low-energy precision tests. We also note that the coupling between the top quark and the $Z$ boson is larger than that between the top quark and the photon, so the sensitivity of $tqZ$ production to new physics is potentially increased compared to $tq\gamma$, which is another advantage relative to the latter process. 

Theoretical calculations for $tqZ$ production are challenging since already at leading order (LO) the processes involve three particles in the final state, with two of them very massive, and four colored particles overall in the hard scattering. The next-to-leading-order (NLO) QCD corrections for $tqZ$ production were calculated in \cite{CER}. The NLO electroweak (EW) corrections as well as off-shell effects were included in \cite{PTV}, with further work in \cite{DPS}. The NLO QCD corrections turn out to be quite significant, providing an enhancement of around 16\% to the LO cross section at LHC energies, while the NLO EW corrections are rather small, only 1\%. Thus, it is important to consider further higher-order QCD corrections. 

Soft-gluon resummation \cite{NKGS1,NKGS2,KOS,NKsingletop,NK2loop,NKsch,NKtW,NKtt2l,NKtch,NK3loop,FK2020} has long been known to be very important for top-quark processes, since the cross section receives large corrections from soft-gluon emission near partonic threshold due to the large mass of the top quark. This is well known for many $2 \to 2$ top-quark processes, including top-antitop pair production \cite{NKGS1,NKGS2,NK2loop,NKtt2l}, single-top production \cite{NKsingletop,NKsch,NKtW,NKtch,NK3loop}, and even processes beyond the Standard Model, for example involving top-quark anomalous couplings \cite{NKAB,NKtZ,NKEM,MGNK} (see e.g. the review in Ref. \cite{NKtoprev}). More recently, soft-gluon resummation has been applied to $2 \to 3$ processes \cite{FK2020}, in particular $tqH$ production \cite{FK2021} and $tq\gamma$ production \cite{NKNY2022}. In all these processes, as well as $tqZ$ production, the soft-gluon corrections are dominant and account for the majority of the complete corrections at NLO. Furthermore, the soft-gluon calculations at next-to-next-to-leading order (NNLO) predicted very well the later complete NNLO results for top-antitop production and $s$-channel single-top production (see e.g. the review in Ref. \cite{NKtoprev}). These facts, along with the relevance of resummation in the related $tZ$ production via anomalous couplings in \cite{NKAB,NKtZ,MGNK}, provide very strong motivation for the study of resummation for $tqZ$ production.

In this paper, we use soft-gluon resummation to calculate approximate NNLO (aNNLO) cross sections for $tqZ$ production. In the next section, we describe the resummation formalism and its specific implementation for $tqZ$ production in single-particle-inclusive kinematics. In Section 3, we provide results for the total cross sections, including theoretical uncertainties, at LHC energies. In Section 4, we give results for the top-quark rapidity distributions. We conclude in Section 5.

\mysection{Resummation for $tqZ$ production}

We begin with the soft-gluon resummation formalism for $tqZ$ production, implementing the theoretical framework in \cite{FK2020}. We study the parton-level processes $a(p_a)+b(p_b) \to t(p_t)+q(p_q)+Z(p_Z)$, and we define the usual kinematical variables $s=(p_a+p_b)^2$, $t=(p_a-p_t)^2$, and $u=(p_b-p_t)^2$. With an additional gluon emission in the final state, momentum conservation is given by $p_a +p_b=p_t +p_q +p_Z+p_g$ where $p_g$ is the gluon momentum. We then define a threshold variable $s_4=(p_q+p_Z+p_g)^2-(p_q+p_Z)^2=s+t+u-m_t^2-(p_q+p_Z)^2$ which involves the extra energy from gluon emission and which vanishes as $p_g \to 0$.

We write the differential cross section for $tqZ$ production in proton-proton collisions as a convolution, 
\beq
d\sigma_{pp \to tqZ}=\sum_{a,b} \; 
\int dx_a \, dx_b \,  \phi_{a/p}(x_a, \mu_F) \, \phi_{b/p}(x_b, \mu_F) \, 
d{\hat \sigma}_{ab \to tqZ}(s_4, \mu_F) \, ,
\label{sigma}
\eeq
where $\mu_F$ is the factorization scale, $\phi_{a/p}$  and $\phi_{b/p}$ are parton distribution functions (pdf) for parton $a$ and parton $b$, respectively, in the proton, and $d{\hat \sigma}_{ab \to tqZ}$ is the partonic differential cross section.

The cross section factorizes if we take Laplace transforms \cite{FK2020}, defined by 
\beq
 d{\tilde{\hat\sigma}}_{ab \to tqZ}(N, \mu_F)=\int_0^s \frac{ds_4}{s} \,  e^{-N s_4/s} \; d{\hat\sigma}_{ab \to tqZ}(s_4, \mu_F), 
\eeq
where $N$ is the transform variable. Under transforms, the logarithms of $s_4$ in the perturbative series go into logarithms of $N$ which, as we will see, exponentiate. We also define transforms of the pdf via ${\tilde \phi}(N)=\int_0^1 e^{-N(1-x)} \phi(x) \, dx$. Replacing the colliding protons by partons in Eq. (\ref{sigma}) \cite{NKGS2,FK2020,GS}, we thus have the factorized form in transform space
\beq
d{\tilde \sigma}_{ab \to tqZ}(N)= {\tilde \phi}_{a/a}(N_a, \mu_F) \, {\tilde \phi}_{b/b}(N_b, \mu_F) \, d{\tilde{\hat \sigma}}_{ab \to tqZ}(N, \mu_F) \, .
\label{factcs}
\eeq

The cross section can be refactorized \cite{NKGS1,NKGS2,KOS,FK2020} in terms of an infrared-safe short-distance hard function, $H_{ab \to tqZ}$,  and a soft function, $S_{ab \to tqZ}$, which describes the emission of noncollinear soft gluons. We have
\beq
d{\tilde{\sigma}}_{ab \to tqZ}(N)={\tilde \psi}_{a/a}(N_a,\mu_F) \, {\tilde \psi}_{b/b}(N_b,\mu_F) \, {\tilde J}_q (N, \mu_F) \, {\rm tr} \left\{H_{ab \to tqZ} \left(\alpha_s(\mu_R)\right) \, {\tilde S}_{ab \to tqZ} \left(\frac{\sqrt{s}}{N \mu_F} \right)\right\} \, ,
\label{refactcs}
\eeq
where $\alpha_s$ is the strong coupling and $\mu_R$ is the renormalization scale. The functions $\psi$ are distributions for incoming partons at fixed value of momentum and involve collinear emission \cite{NKGS1,NKGS2,KOS,GS} while the function $J_q$ describes radiation from the final-state light quark. The hard and the soft functions for the $tqZ$ production processes are $2\times 2$ matrices in the color space of the partonic scattering.

Comparing Eqs. (\ref{factcs}) and (\ref{refactcs}), we find an expression for the hard-scattering partonic cross section in transform space
\beq
d{\tilde{\hat \sigma}}_{ab \to tqZ}(N, \mu_F)=
\frac{{\tilde \psi}_{a/a}(N_a, \mu_F) \, {\tilde \psi}_{b/b}(N_b, \mu_F) \, {\tilde J_q} (N, \mu_F)}{{\tilde \phi}_{a/a}(N_a, \mu_F) \, {\tilde \phi}_{b/b}(N_b, \mu_F)} \; \,  {\rm tr} \left\{H_{ab \to tqZ}\left(\alpha_s(\mu_R)\right) \, 
{\tilde S}_{ab \to tqZ}\left(\frac{\sqrt{s}}{N \mu_F} \right)\right\} \, .
\label{sigN}
\eeq

The dependence of the soft matrix on the transform variable, $N$, is resummed via renormalization-group evolution \cite{NKGS1,NKGS2}. Thus, ${\tilde S}_{ab \to tqZ}$ obeys a renormalization-group equation in terms of a soft anomalous dimension matrix, $\Gamma_{\! S \, ab \to tqZ}$, which is calculated from the coefficients of the ultraviolet poles of the relevant eikonal diagrams \cite{NKGS1,NKGS2,KOS,FK2020,NKsingletop,NKsch,NKtW,NKtch,NK2loop,NKtt2l,NK3loop,NKtoprev}.

The $N$-space resummed cross section, which resums logarithms of $N$, is derived from the renormalization-group evolution of the functions ${\tilde S}_{ab \to tqZ}$, ${\tilde \psi}$, ${\tilde \phi}$, and ${\tilde J}_q$ in Eq. (\ref{sigN}), and it is given by
\beqa
d{\tilde{\hat \sigma}}_{ab \to tqZ}^{\rm resum}(N,\mu_F) &=&
\exp\left[\sum_{i=a,b} E_{i}(N_i)\right] \, 
\exp\left[\sum_{i=a,b} 2 \int_{\mu_F}^{\sqrt{s}} \frac{d\mu}{\mu} \gamma_{i/i}(N_i)\right] \, 
\exp\left[E'_q(N)\right]
\nonumber\\ && \hspace{-5mm} \times \,
{\rm tr} \left\{H_{ab \to tqZ}\left(\alpha_s(\sqrt{s})\right) {\bar P} \exp \left[\int_{\sqrt{s}}^{{\sqrt{s}}/N}
\frac{d\mu}{\mu} \; \Gamma_{\! S \, ab \to tqZ}^{\dagger} \left(\alpha_s(\mu)\right)\right] \; \right.
\nonumber\\ && \left. \hspace{5mm} \times \,
{\tilde S}_{ab \to tqZ} \left(\alpha_s\left(\frac{\sqrt{s}}{N}\right)\right) \;
P \exp \left[\int_{\sqrt{s}}^{{\sqrt{s}}/N}
\frac{d\mu}{\mu}\; \Gamma_{\! S \, ab \to tqZ}
\left(\alpha_s(\mu)\right)\right] \right\} \, ,
\nonumber \\
\label{resummed}
\eeqa
where $P$ (${\bar P}$) denotes path-ordering in the same (reverse) sense as the integration variable $\mu$.
The first exponential in Eq. (\ref{resummed}) resums soft and collinear emission from the initial-state partons, while the second exponential involves the parton anomalous dimensions $\gamma_{i/i}$ and the dependence on $\mu_F$. The third exponential resums radiation from the final-state light quark. The last two exponentials involve integrals of the soft anomalous dimension $\Gamma_{\! S \, ab \to tqZ}$ and its Hermitian adjoint. Explicit results for all the functions in the exponents of Eq. (\ref{resummed}) at one and two loops can be found in Ref. \cite{FK2020}. The resummed cross section can be expanded at fixed order and then inverted to momentum space, without requiring a prescription, to produce numerical results.

\mysection{Total $tqZ$ cross sections}

In this section we present results for the total cross sections for $tqZ$ production and, separately, for  ${\bar t}qZ$ production (i.e. with an antitop). We set the top-quark mass $m_t=172.5$ GeV, and the factorization and renormalization scales equal to each other, with this common scale denoted by $\mu$. The NLO results are found by using {\small \sc MadGraph5\_aMC@NLO} \cite{MG5}. We use MSHT20 \cite{MSHT20} pdf in our calculations unless otherwise noted.

We first note that the soft-gluon corrections at NLO provide an excellent approximation to the exact NLO results. If we define the approximate NLO (aNLO) cross section as the sum of the LO cross section and the first-order soft-gluon corrections, then we find that the aNLO result is very close to the complete NLO one for LHC energies, with less than 1\% difference, indicating that the overwhelming majority of the NLO corrections are accounted for in our formalism. We note that this is very similar to the findings for the related $tZ$ production process via anomalous couplings \cite{NKtZ,MGNK}. This provides very strong motivation and justification for the calculation of aNNLO corrections.

Our aNNLO results are calculated by adding the second-order soft-gluon corrections to the exact NLO result; i.e., aNNLO=NLO+ soft aNNLO corrections. These aNNLO calculations provide our best theoretical prediction for the cross section.

\begin{figure}[htbp]
\begin{center}
\includegraphics[width=88mm]{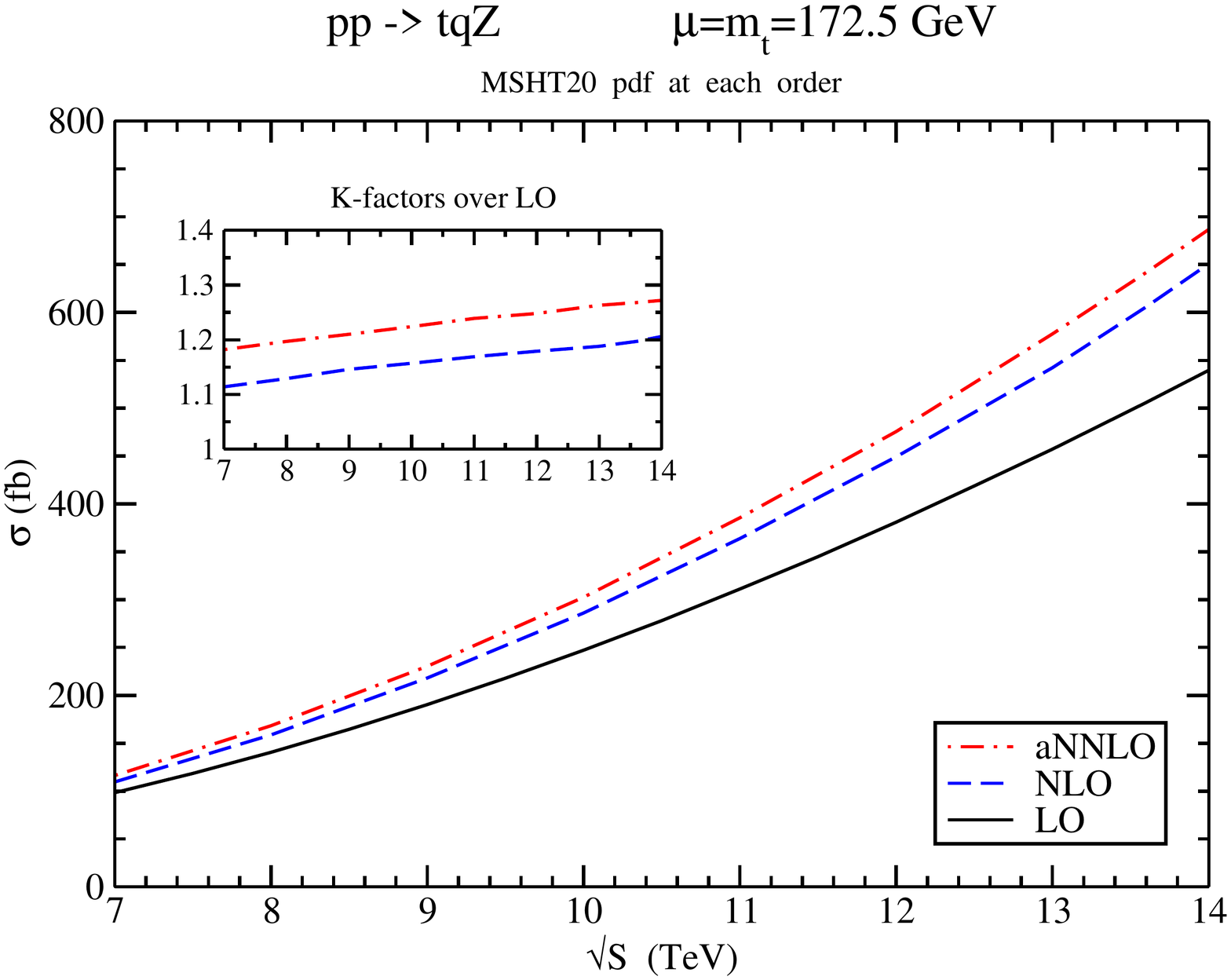}
\includegraphics[width=88mm]{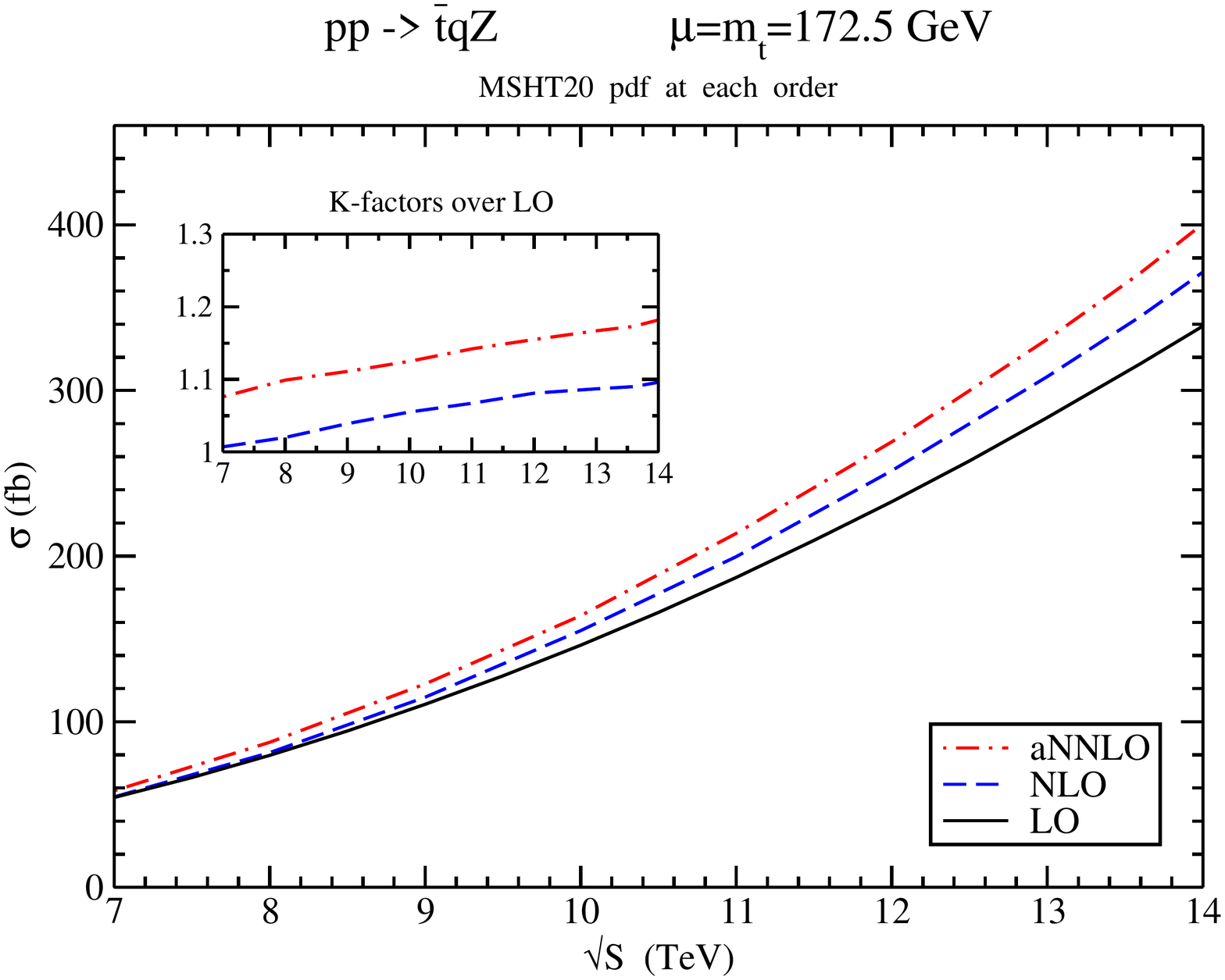}
\caption{The total cross sections for $tqZ$ (left) and ${\bar t}qZ$ (right) production in $pp$ collisions at LHC energies.}
\label{tqZ}
\end{center}
\end{figure}

In Figure \ref{tqZ} we show results for the $tqZ$ (left plot) and ${\bar t}qZ$ (right plot) cross sections at LO, NLO, and aNNLO in $pp$ collisions at LHC energies using MSHT20 pdf corresponding at each order. The inset plots show the $K$-factors relative to LO, i.e. the NLO/LO and the aNNLO/LO ratios. We see that the $K$-factors gradually increase as the energy goes from 7 to 14 TeV.

\begin{table}[htbp]
\begin{center}
\begin{tabular}{|c|c|c|c|c|c|c|c|c|} \hline
\multicolumn{6}{|c|}{$tqZ$ cross sections in $pp$ collisions at the LHC} \\ \hline
$\sigma$ in fb & 7 TeV & 8 TeV & 13 TeV & 13.6 TeV & 14 TeV \\ \hline
LO    & $98.4{}^{+1.6}_{-3.9} {}^{+1.6}_{-0.9}$ & $141^{+3}_{-7} \pm 2$ & $457^{+21}_{-32}{}^{+6}_{-3}$ & $506^{+24}_{-36}{}^{+6}_{-4}$ & $540^{+26}_{-40}{}^{+6}_{-4}$ \\ \hline
NLO   & $110 \pm 2 {}^{+1}_{-2}$  & $159^{+3}_{-2} \pm 2$ & $542 \pm 11 {}^{+6}_{-5}$ & $606^{+11}_{-12} \pm 6$ & $651^{+14}_{-16} \pm 6$ \\ \hline
aNNLO & $116{}^{+1}_{-2} \pm 2$  & $168 \pm 2 {}^{+3}_{-2}$ & $577^{+4}_{-9}{}^{+6}_{-5}$ & $641^{+4}_{-10}{}^{+6}_{-5}$ & $686^{+5}_{-13}{}^{+7}_{-5}$ \\ \hline
\end{tabular}
\caption[]{The $tqZ$ cross sections (in fb), with scale and pdf uncertainties, in $pp$ collisions with $\sqrt{S}=7$, 8, 13, 13.6, and 14 TeV, $m_t=172.5$ GeV, and MSHT20 pdf.}
\label{table1}
\end{center}
\end{table}

In Table 1 we show total rates for $tqZ$ production for various LHC energies at LO, NLO, and aNNLO using MSHT20 pdf at each order. The central values are with a scale choice $\mu=m_t=172.5$ GeV, the first uncertainty is from scale variation over the range $m_t/2$ to $2m_t$, and the second uncertainty is from the MSHT20 pdf. The NLO corrections increase the LO cross section by 12\% at 7 TeV, 13\% at 8 TeV, 19\% at 13 TeV, 20\% at 13.6 TeV, and 21\% at 14 TeV. The aNNLO corrections are also significant, providing a further increase of around 6\% at 7 and 8 TeV, and around 7\% at 13, 13.6, and 14 TeV. Furthermore, the scale dependence is reduced significantly at higher orders, and at aNNLO it becomes comparable to the pdf uncertainty. Thus, the aNNLO result provides a significantly improved theoretical prediction. 

We note that the use of other recent pdf sets, i.e.  CT18 pdf \cite{CT18} and NNPDF4.0 pdf \cite{NNPDF4.0}, gives NLO cross sections (with NLO pdf) that are almost identical to the ones we derived with MSHT20 pdf in both central value and scale uncertainty. For example, the NLO cross section at 13 TeV with CT18 pdf is $541 \pm 11 {}^{+14}_{-12}$ fb, and with NNPDF4.0 pdf it is $538 \pm 11 \pm 3$ fb, while the corresponding result with MSHT20 pdf, as appears in Table 1, is $542 \pm 11 {}^{+6}_{-5}$ fb.
At aNNLO (with NNLO pdf), the cross section with CT18 pdf is $582^{+4}_{-9} {}^{+14}_{-13}$ fb, and with NNPDF4.0 pdf it is $560^{+4}_{-9} \pm 2$ fb, while the corresponding result with MSHT20 pdf, as given in Table 1, is $577^{+4}_{-9} {}^{+6}_{-5}$ fb.

\begin{table}[htbp]
\begin{center}
\begin{tabular}{|c|c|c|c|c|c|c|c|c|} \hline
\multicolumn{6}{|c|}{${\bar t}qZ$ cross sections in $pp$ collisions at the LHC} \\ \hline
$\sigma$ in fb & 7 TeV & 8 TeV & 13 TeV & 13.6 TeV & 14 TeV \\ \hline
LO    & $54.3^{+0.8}_{-2.1}{}^{+0.9}_{-1.2}$ & $79.7^{+1.8}_{-3.6}{}^{+1.2}_{-1.5}$ & $284^{+13}_{-20}{}^{+3}_{-5}$ & $316^{+15}_{-23}{}^{+3}_{-5}$ & $339^{+17}_{-25}{}^{+3}_{-5}$ \\ \hline
NLO   & $54.6^{+1.3}_{-0.8}{}^{+1.3}_{-1.0}$ & $81.3^{+1.9}_{-1.3}{}^{+1.7}_{-1.4}$ & $308^{+8}_{-7}{}^{+5}_{-4}$ & $345^{+8}_{-9} \pm 5$ & $371^{+10}_{-9}{}^{+6}_{-4}$ \\ \hline
aNNLO & $58.4^{+0.4}_{-0.7}{}^{+1.2}_{-1.0}$ & $87.6^{+0.4}_{-1.2}{}^{+1.7}_{-1.4}$ & $331^{+2}_{-6} \pm 4$ & $371^{+2}_{-8} \pm 4$ & $401^{+2}_{-8}{}^{+5}_{-4}$ \\ \hline
\end{tabular}
\caption[]{The ${\bar t}qZ$ cross sections (in fb), with scale and pdf uncertainties, in $pp$ collisions with $\sqrt{S}=7$, 8, 13, 13.6, and 14 TeV, $m_t=172.5$ GeV, and MSHT20 pdf.}
\label{table2}
\end{center}
\end{table}

In Table 2 we show total rates for ${\bar t}qZ$ production (i.e. with an antitop) for various LHC energies at LO, NLO, and aNNLO using MSHT20 pdf at each order. In this case, the NLO corrections are significantly smaller than for $tqZ$ production while the aNNLO corrections are comparatively large, and one might worry about the convergence of the perturbative series. However, much of the difference between the aNNLO and NLO results in ${\bar t}qZ$ production is due to the differences in the pdf used at each order. For example, at 13.6 TeV, using MSHT20 NNLO pdf for all orders, the ${\bar t}qZ$ cross section is 320 fb at LO, 357 fb at NLO, and 371 fb at aNNLO. Also, the $tqZ$ cross section at 13.6 TeV with MSHT20 NNLO pdf is 583 fb at LO, 627 fb at NLO, and 641 fb at aNNLO.

\mysection{Top-quark rapidity distributions}

Next, we present the top-quark rapidity ($y_t$) distributions in $tqZ$ production, which provide more information than total cross sections. We present results at 13, 13.6, and 14 TeV energies.

\begin{figure}[htbp]
\begin{center}
\includegraphics[width=88mm]{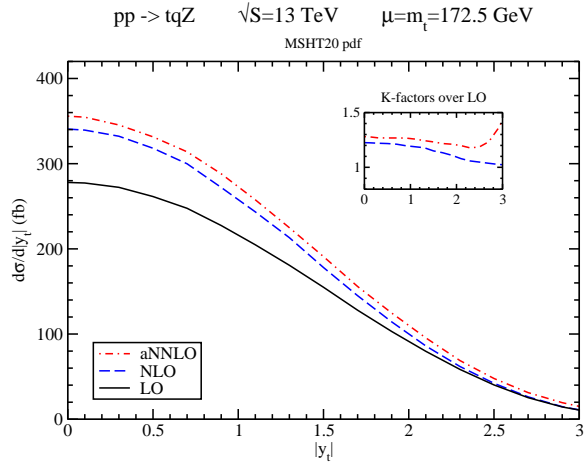}
\caption{The top-quark rapidity distributions in $tqZ$ production at 13 TeV LHC energy.}
\label{yabstoptqZ13}
\end{center}
\end{figure}

Figure \ref{yabstoptqZ13} shows the top-quark rapidity distributions in $tqZ$ production at 13 TeV LHC energy. The LO, NLO, and aNNLO results are plotted for the distribution of the absolute value of the top-quark rapidity, $d\sigma/d|y_t|$. The $K$-factors are shown in the inset plot. The aNNLO/LO line clearly increases at very high values of the top-quark rapidity. As we discussed for the total cross sections, part of the difference between the aNNLO and NLO results is due to the different pdf, especially at large rapidities.

\begin{figure}[htbp]
\begin{center}
\includegraphics[width=88mm]{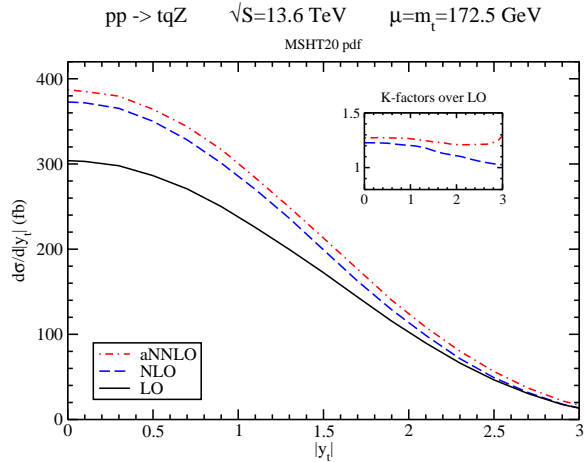}
\caption{The top-quark rapidity distributions in $tqZ$ production at 13.6 TeV LHC energy.}
\label{yabstoptqZ13.6}
\end{center}
\end{figure}

Figure \ref{yabstoptqZ13.6} shows the LO, NLO, and aNNLO top-quark rapidity distributions in $tqZ$ production at 13.6 TeV LHC energy, with the $K$-factors in the inset plot. Again, we observe significant enhancements from the aNNLO crorrections, especially at large rapidities. 

\begin{figure}[htbp]
\begin{center}
\includegraphics[width=88mm]{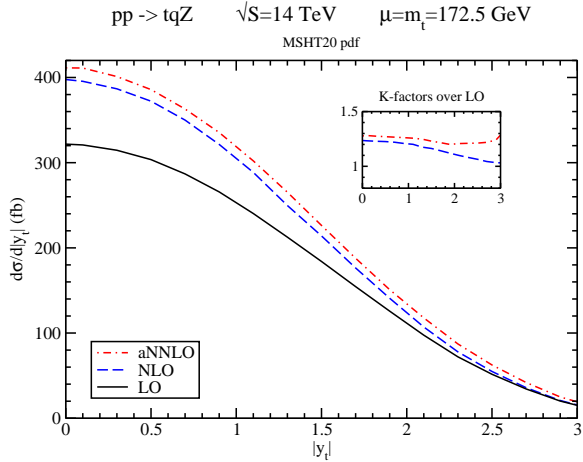}
\caption{The top-quark rapidity distributions in $tqZ$ production at 14 TeV LHC energy.}
\label{yabstoptqZ14}
\end{center}
\end{figure}

Figure \ref{yabstoptqZ14} shows the LO, NLO, and aNNLO top-quark rapidity distributions in $tqZ$ production at 14 TeV LHC energy. The $K$-factors in the inset plot show, again, the importance of the aNNLO corrections, particularly at large $y_t$.

Finally, we discuss the scale and pdf uncertainties in the rapidity distributions. For central and moderate values of the rapidity, up to a rapidity value of 2, the fractional scale and pdf uncertainties of the rapidity distribution at each order are essentially the same as those for the corresponding total cross section. However, for large rapidity values both the scale and the pdf uncertainties get bigger. At a rapidity value of 2.5, the fractional scale uncertainty is double that of the total cross section, while at a rapidity value of 3 it is triple. Also, the fractional pdf uncertainty is three times bigger than that of the total cross section at a rapidity value of 2.5 and six times bigger at a rapidity value of 3.

\mysection{Conclusions}

We have calculated higher-order QCD corrections, and more specifically soft-gluon corrections, for the associated production of a single top quark with a $Z$ boson, which is a process that has been actively studied at the LHC. The NLO QCD corrections are very significant and, thus, it is important to consider even higher-order corrections. By comparing the soft-gluon corrections at NLO with the complete NLO result, we have found that the corrections are dominated by soft-gluon emission. Thus, the calculation of aNNLO cross sections, i.e. the complete NLO result plus second-order soft-gluon corrections, is an important step towards producing better theoretical predictions.

We have found that the aNNLO corrections, which are derived from soft-gluon resummation, are very significant and increase the NLO cross section at LHC energies. We have also calculated theoretical uncertainties from scale variation and from pdf. We have found that the scale uncertainties decrease with each higher perturbative order. The pdf uncertainties remain relatively small at all orders.

Finally, we have calculated the rapidity distributions of the top quark at LO, NLO, and aNNLO. There are significant enhancements from the aNNLO corrections, particularly at larger rapidity values.

Our results provide the highest-order and up-to-date theoretical predictions for $tqZ$ production at past, current, and future LHC energies. The improved precision is important as this process is sensitive to various couplings of the top quark and the $W$ and $Z$ bosons, and it can probe anomalous weak moments of the top quark.

\section*{Acknowledgements}
We thank Matthew Forslund for early contributions to this work.
The work of N.K. is supported by the National Science Foundation under Grant No. PHY 2112025.


\begin{thebibliography}{99}

\bibitem{CMS1}
CMS Collaboration, {\sl Search for associated production of a $Z$ boson with a single top quark and for $tZ$ flavour-changing interactions in $pp$ collisions at $\sqrt{s}=8$ TeV}, JHEP {\bf 07}, 003 (2017) [arXiv:1702.01404].

\bibitem{ATLAS1}
ATLAS Collaboration, {\sl Measurement of the production cross-section of a single top quark in association with a $Z$ boson in proton-proton collisions at 13 TeV with the ATLAS detector}, Phys. Lett. B {\bf 780}, 557 (2018) [arXiv:1710.03659].

\bibitem{CMS2}
CMS Collaboration, {\sl Measurement of the associated production of a single top quark and a $Z$ boson in $pp$ collisions at $\sqrt{s}= 13$ TeV}, Phys. Lett. B {\bf 779}, 358 (2018) [arXiv:1712.02825].  

\bibitem{CMS3}
CMS Collaboration, {\sl Observation of Single Top Quark Production in Association with a $Z$ Boson in Proton-Proton Collisions at $\sqrt{s}=13$ TeV}, Phys. Rev. Lett. {\bf 122}, 132003 (2019) [arXiv:1812.05900].

\bibitem{ATLAS2}
ATLAS Collaboration, {\sl Observation of the associated production of a top quark and a $Z$ boson in $pp$ collisions at $\sqrt{s}=13$ TeV with the ATLAS detector}, JHEP {\bf 07}, 124 (2020) [arXiv:2002.07546].

\bibitem{CMS4}
CMS Collaboration, {\sl Inclusive and differential cross section measurements of single top quark production in association with a $Z$ boson in proton-proton collisions at $\sqrt{s}= 13$ TeV}, JHEP {\bf 02}, 107 (2022) [arXiv:2111.02860].

\bibitem{NKAB}
N. Kidonakis and A. Belyaev, {\sl FCNC top quark production via anomalous $tqV$  couplings beyond leading order}, JHEP {\bf 12}, 004 (2003) [arXiv:hep-ph/0310299].

\bibitem{LZLGZ}
B.H. Li, Y. Zhang, C.S. Li, J. Gao, and H. X. Zhu, {\sl Next-to-leading order QCD corrections to $tZ$ associated production via the flavor-changing neutral-current couplings at hadron colliders}, Phys. Rev. D {\bf 83}, 114049 (2011) [arXiv:1103.5122].

\bibitem{NKtZ}
N. Kidonakis, {\sl Higher-order corrections for $tZ$ production via anomalous couplings}, Phys. Rev. D {\bf 97}, 034028 (2018) [arXiv:1712.01144].

\bibitem{MGNK}
M. Guzzi and N. Kidonakis, {\sl $tZ'$ production at hadron colliders}, Eur. Phys. J. C {\bf 80}, 467 (2020) [arXiv:1904.10071].

\bibitem{YLSM}
Y.-B. Liu and S. Moretti, {\sl Probing $tqZ$ anomalous couplings in the trilepton signal at the HL-LHC, HE-LHC and FCC-hh}, Chin. Phys. C {\bf 45}, 043110 (2021) [arXiv:2010.05148].

\bibitem{NKEM}
N. Kidonakis and E. Martin, {\sl Soft-Gluon Corrections in FCNC Top-Quark Production via Anomalous Gluon Couplings}, Phys. Rev. D {\bf 90}, 054021 (2014) [arXiv:1404.7488]. 

\bibitem{JRMT}
J. Reuter and M. Tonini, {\sl Top Partner Discovery in the $T \to tZ$ channel at the LHC}, JHEP {\bf 01}, 088 (2015) [arXiv:1409.6962].

\bibitem{DMMVZ}
C. Degrande, F. Maltoni, K. Mimasu, E. Vryonidou, and C. Zhang, {\sl Single-top associated production with a $Z$ or $H$ boson at the LHC: the SMEFT interpretation}, JHEP {\bf 10}, 005 (2018) [arXiv:1804.07773].

\bibitem{RBAI}
R.K. Barman and A. Ismail, {\sl Constraining the top electroweak sector of the SMEFT through $Z$ associated top pair and single top production at the HL-LHC}, arXiv:2205.07912.

\bibitem{CDVM}
V. Cirigliano, W. Dekens, J. de Vries and E. Mereghetti, {\sl Is there room for CP violation in the top-Higgs sector?}, Phys. Rev. D {\bf 94}, 016002 (2016) [arXiv:1603.03049].

\bibitem{Yamanaka2017}
N. Yamanaka, B.K. Sahoo, N. Yoshinaga, T. Sato, K. Asahi and B.P. Das,
{\sl Probing exotic phenomena at the interface of nuclear and particle physics with the electric dipole moments of diamagnetic atoms: A unique window to hadronic and semi-leptonic CP violation}, Eur. Phys. J. A {\bf 53}, 54 (2017) [arXiv:1703.01570].

\bibitem{CER}
J. Campbell, R.K. Ellis, and R. Rontsch, {\sl Single top production in association with a $Z$ boson at the LHC}, Phys. Rev. D {\bf 87}, 114006 (2013) [arXiv:1302.3856].

\bibitem{PTV}
D. Pagani, I. Tsinikos, and E. Vryonidou, {\sl NLO QCD+EW predictions for $tHj$ and $tZj$ production at the LHC}, JHEP {\bf 08}, 082 (2020) [arXiv:2006.10086].

\bibitem{DPS}
A. Denner, G. Pelliccioli, and C. Schwan, {\sl NLO QCD and EW corrections to off-shell $tZj$ production at the LHC}, arXiv:2207.11264.

\bibitem{NKGS1}
N. Kidonakis and G. Sterman, {\sl Subleading logarithms in QCD hard scattering}, Phys. Lett. B {\bf 387}, 867 (1996). 

\bibitem{NKGS2}
N. Kidonakis and G. Sterman, {\sl Resummation for QCD hard scattering}, 
Nucl. Phys. B {\bf 505}, 321 (1997) [hep-ph/9705234].

\bibitem{KOS}
N. Kidonakis, G. Oderda, and G. Sterman, {\sl Evolution of color exchange in QCD hard scattering}, Nucl. Phys. B {\bf 531}, 365 (1998) [hep-ph/9803241].

\bibitem{NKsingletop}
N. Kidonakis, {\sl Single top production at the Tevatron: threshold resummation and finite-order soft gluon corrections}, Phys. Rev. D {\bf 74}, 114012 (2006) [arXiv:hep-ph/0609287]. 

\bibitem{NK2loop}
N. Kidonakis, {\sl Two-loop soft anomalous dimensions and NNLL resummation for heavy quark production}, Phys. Rev. Lett. {\bf 102}, 232003 (2009) [arXiv:0903.2561].

\bibitem{NKsch}
N. Kidonakis, {\sl NNLL resummation for $s$-channel single top quark production}, Phys. Rev. D {\bf 81}, 054028 (2010) [arXiv:1001.5034].

\bibitem{NKtW}
N. Kidonakis, {\sl Two-loop soft anomalous dimensions for single top quark associated production with a $W^-$ or $H^-$}, Phys. Rev. D {\bf 82}, 054018 (2010) [arXiv:1005.4451]. 

\bibitem{NKtt2l}
N. Kidonakis, {\sl Next-to-next-to-leading soft-gluon corrections for the top quark cross section and transverse momentum distribution}, Phys. Rev. D {\bf 82}, 114030 (2010) [arXiv:1009.4935].

\bibitem{NKtch}
N. Kidonakis, {\sl Next-to-next-to-leading-order collinear and soft gluon corrections for $t$-channel single top quark production}, Phys. Rev. D {\bf 83}, 091503 (2011) [arXiv:1103.2792]. 

\bibitem{NK3loop}
N. Kidonakis, {\sl Soft anomalous dimensions for single-top production at three loops}, Phys. Rev. D {\bf 99}, 074024 (2019) [arXiv:1901.09928].

\bibitem{FK2020}
M. Forslund and N. Kidonakis, {\sl Resummation for $2 \to n$ processes in single-particle-inclusive kinematics}, Phys. Rev. D {\bf 102}, 034006 (2020) [arXiv:2003.09021].

\bibitem{NKtoprev}
N. Kidonakis, {\sl Soft-gluon corrections in top-quark production}, Int. J. Mod. Phys. A {\bf 33}, 1830021 (2018) [arXiv:1806.03336].

\bibitem{FK2021}
M. Forslund and N. Kidonakis, {\sl Soft-gluon corrections for the associated production of a single top quark and a Higgs boson}, Phys. Rev. D {\bf 104}, 034024 (2021) [arXiv:2103.01228].

\bibitem{NKNY2022}
N. Kidonakis and N. Yamanaka, {\sl QCD corrections in $tq\gamma$ production at hadron colliders}, Eur. Phys. J. C {\bf 82}, 670 (2022) [arXiv:2201.12877].

\bibitem{GS}
G. Sterman, {\sl Summation of large corrections to short-distance hadronic cross sections}, Nucl. Phys. B {\bf 281}, 310 (1987). 

\bibitem{MG5}
J. Alwall {\it et al.}, {\sl The automated computation of tree-level and next-to-leading order differential cross sections, and their matching to parton shower simulations}, JHEP {\bf 07}, 079 (2014) [arXiv:1405.0301].

\bibitem{MSHT20}
S. Bailey, T. Cridge, L.A. Harland-Lang, A.D. Martin, and R.S. Thorne, {\sl Parton distributions from LHC, HERA, Tevatron and fixed target data: MSHT20 PDFs}, Eur. Phys. J. C {\bf 81}, 341 (2021) [arXiv:2012.04684].

\bibitem{CT18}
T.-J. Hou {\it et al.}, {\sl New CTEQ global analysis of quantum chromodynamics with high-precision data from the LHC}, Phys. Rev. D {\bf 103}, 014013 (2021) [arXiv: 1912.10053].

\bibitem{NNPDF4.0}
R.D. Ball {\it et al.}, {\sl The path to proton structure at one-percent accuracy}, Eur. Phys. J. C {\bf 82}, 428 (2022) [arXiv:2109.02653].

\end{thebibliography}
\end{document}